\documentclass[aps,pra,preprint,groupedaddress, showpacs]{revtex4}

\usepackage{graphicx}
\pdfoutput=1

\begin{document}


\title{X-ray emission measurements following charge exchange between C$^{6+}$ and H$_2$}


\author{M.~Fogle}
\email[]{fogle@physics.auburn.edu}
\affiliation{Department of Physics, Auburn University, Auburn, Alabama 36849}

\author{D.~Wulf}
\affiliation{Department of Physics, University of Wisconsin, Madison, Wisconsin 53706}

\author{K.~Morgan}
\affiliation{Department of Physics, University of Wisconsin, Madison, Wisconsin 53706}

\author{D.~McCammon}
\affiliation{Department of Physics, University of Wisconsin, Madison, Wisconsin 53706}

\author{D.~G.~Seely}
\affiliation{Department of Physics, Albion College, Albion, Michigan}



\author{I.~N.~Dragani\'c}
\affiliation{Physics Division, Oak Ridge National Laboratory, Oak Ridge, Tennessee 37831}

\author{C.~C.~Havener}
\affiliation{Physics Division, Oak Ridge National Laboratory, Oak Ridge, Tennessee 37831}


\date{\today}

\begin{abstract}
Lyman x-ray spectra following charge exchange between C$^{6+}$ and H$_2$ are presented for collision velocities between 400 and 2300~km/s (1--30~keV/amu). Spectra were measured by a microcalorimeter x-ray detector capable of fully resolving the C VI Lyman series emission lines though Lyman-$\delta$. The ratios of the measured emission lines are sensitive to the angular momentum $l$-states populated during charge exchange and are used to gauge the effectiveness of different $l$-distribution models in predicting Lyman emission due to charge exchange. At low velocities, we observe that both single electron capture and double capture autoionization contribute to Lyman emission and that a statistical $l$-distribution best describes the measured line ratios. At higher velocities single electron capture dominates with the $l$-distribution peaked at the maximum $l$.
\end{abstract}


\maketitle

\section{Introduction}
Charge exchange (CX) between highly charged ions and atomic and molecular targets can exhibit large cross sections ($10^{-15}$~cm$^2$) at certain collision velocity regimes, making it one of the dominant processes in plasma environments. The resulting emission lines due to cascading captured electrons can provide temperature, density and relative abundance information about the interaction environment. In laboratory magnetic confinement plasmas, CX is a common diagnostic tool used in combination with puffed gas and neutral beams injection \cite{hulse1980,mattioli1989,isler1994}. In astrophysical settings, the identification of x-ray emission from comets has been linked to CX between solar wind ions and ablated cometary neutral gases \cite{lisse1996,cravens1997, beiersdorfer2001}. These same solar wind ions interact with neutrals of planetary atmospheres and from the heliosphere. These applications have made CX modeling of paramount interest to both the laboratory plasma and astrophysics communities. 

A successful model of emission spectra due to CX relies on being able to map the specific distribution of principal, $n$, and angular momentum, $l$, quantum states in which the transferred electron is captured on the ion. Several successful tools have been developed to estimate the $n$-state of capture, however, the state-selective $n,l$ CX cross sections are highly dependent on collision velocity and several models, of various effectiveness, have been put forth to estimate the $l$-distributions in CX over different collision velocity regimes.

A number of previous theoretical and experimental studies helped to establish methods of determining CX cross sections and the principle quantum state $n$ of capture for given interaction pairs and energies. In the regime where the collision velocity is approximately equal to the orbital velocity of the captured electron, the most successful tool has been the classical over-the-barrier model (CBM) \cite{ryufuku1980}. At higher collision velocities, the classical trajectory Monte Carlo (CTMC) technique has shown to be successful \cite{simcic2010}. At lower collision velocities, the CX process involves complex trajectories and interaction dynamics between many states and is best modeled by atomic orbital (or molecular orbital) close-coupling (AOCC, MOCC) methods \cite{fritsch1986, kimura1986b}. Given that these calculations involve multiple electrons (and nuclei) they are inherently difficult and have not been widely adapted for extended use in CX modeling. It should also be noted that at these slower collision velocities, transfer ionization (TI), stabilized double capture (DC) and double capture autoionization (DCAI) can contribute to the total CX cross section in addition to the normal single electron capture (SEC) mechanism. These additional processes are rarely included in CX models.

The most prolific $l$-distribution models from the literature \cite{smith2012, janev1985} have been $i$) an even distribution, in which the $l$-states are evenly distributed across an $n$-manifold based on the total number of angular momentum states available, $ii$) a statistical distribution given by
\begin{equation}
\frac{2l+1}{n^2} \; \mbox{,}
\end{equation}
 $iii$) a Landau-Zener (LZ) distribution given by
\begin{equation}
\frac{l(l+1)(2l+1)(n-1)!(n-2)!}{(n+l)!(n-l-1)!}
\end{equation}
and $iv$) a separable distribution given by
\begin{equation}
\frac{2l+1}{Z}\exp\left( \frac{-l(l+1)}{Z}\right)
\end{equation}
where $Z$ is the charge of the capturing ion. In general, a statistical distribution is usually assumed to become dominate when the collision velocity is approximately half the orbital velocity of the captured electron \cite{burgdorfer1986}. At slower collision velocities the $l$-distribution tends to be either even or peaked at intermediate $l$-states, such as in the separable and LZ distributions above. At higher collision velocities, the $l$-distribution migrates to preferentially higher $l$-states that can lead to \emph{over-statistical} distributions with a majority of captured electrons in the maximum $l$-state \cite{ryufuku1979, olson1981, green1982, dijkkamp1984, fritsch1984, kimura1985, toshima1995}. None of the $l$-distribution models above describe this over-statistical distribution condition.

Here we report the measurement of Lyman emission lines resulting from the CX interaction between C$^{6+}$ and H$_2$ using a high resolution microcalorimeter x-ray detector. From these measured emission lines, we investigate the effect of collision velocity on the state-selective $l$-distributions during the capture process by determining line ratios between Lyman-alpha (L$_\alpha$), Lyman-beta (L$_\beta$) and Lyman-gamma (L$_\gamma$) emission lines.

Total CX cross sections for C$^{6+}$ on H$_2$ and He have been measured by Meyer et al. \cite{meyer1985, meyer1985b} and Greenwood et al. \cite{greenwood2001}. Few state-resolved CX experiments, however, have been conducted for C$^{6+}$ on H$_2$. Dijkkamp et al. \cite{dijkkamp1985} utilized vacuum ultraviolet (VUV) spectroscopy to investigate the state-resolved $n,l$ cross sections of CX for C$^{6+}$, N$^{6+}$ and O$^{6+}$ on He and H$_2$, however, their results for C$^{6+}$ on H$_2$ were inconclusive due to high relative uncertainties and unavoidable state degeneracies related the the observed emission lines. Hoekstra et al. \cite{hoekstra1989} repeated the VUV measurements of Dijkkamp et al. with a refined experimental apparatus with specific aim to investigate the relative $n$ capture states via SEC and DCAI. Mack et al. \cite{mack1987,mack1989} measured correlation effects of double capture between C$^{6+}$ and H$_2$. Recently, we reported high resolution Lyman emission measurements resulting from CX between C$^{6+}$ and He using the microcalorimeter apparatus discussed here \cite{defay2013}. All of these previous results clearly show that even though H$_2$ and He are both strict two-electron systems, the ionization potential of the target is the primary parameter that governs the $n$-state capture. For the He target, the principal capture in C$^{6+}$ is to $n=3$, which is confirmed by the CBM. For H$_2$, the principal capture state is $n=4$, also confirmed by the CBM. In comparing our measured Lyman emission lines between CX with H$_2$ and He we observe significantly different line ratios that suggest different $l$-distributions over the same collision velocities. The particular $l$-distributions, however, are not so easily estimated at different collision velocity regimes. It is the intent here to provide further information on the velocity dependence of CX between C$^{6+}$ and H$_2$ with regard to the capture $l$-states. Given that no detailed theory exists for this collision system over this energy range, we are utilizing the $l$-distribution models above to help describe the observed line ratios. These models are the most widely used by laboratory and astrophysical plasma modelers in accounting for charge exchange in the absence of rigorous theory. These models, while based on agreement with other collision systems, are approximations and are not a rigorous replacement for detailed theoretical calculations. For example, these $l$-distribution models do not account for the quantum defect of $s$-states which will vary based on principle quantum number and will impact the overall CX dynamics. To that end, we will present each of these models in comparison to our measured line ratios as a function of collision velocity. We will also incorporate the overall distribution of the relative $n$-state capture cross sections to account for SEC and DCAI.


\section{Experiment}
\begin{figure}
\includegraphics[scale=0.6]{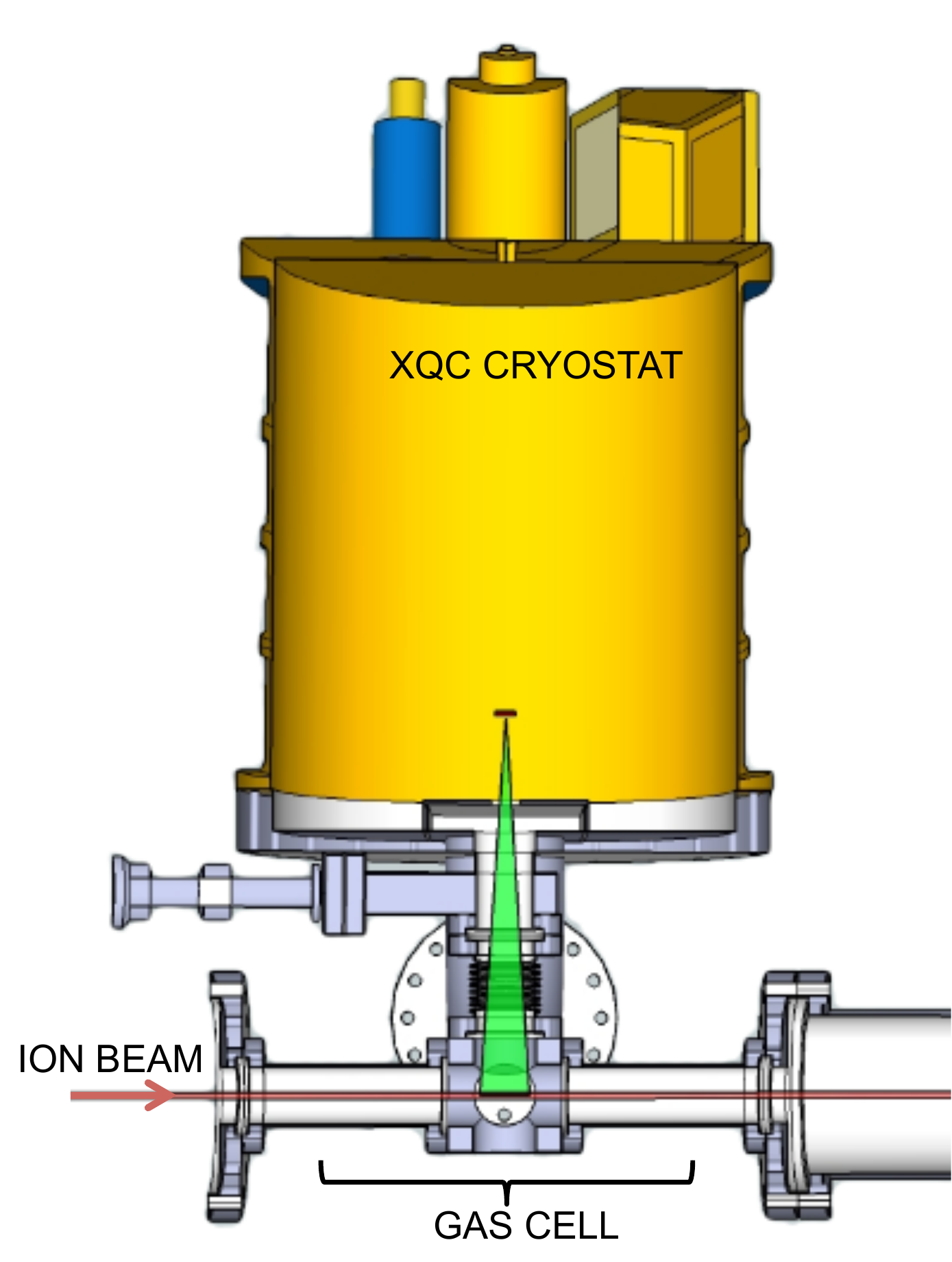}
\caption{(color online) Schematic of the charge exchange cell with XQC. The viewable portion of the gas cell is shown by the viewing angle from the detector array.\label{detectorfig}}
\end{figure}

The charge exchange emission spectra of C$^{6+}$ on H$_2$ were measured by adapting the ion-atom merged beams apparatus at Oak Ridge National Laboratory \cite{draganic2011} with an X-ray microcalorimeter detector from the University of Wisconsin and Goddard Space Flight Center sounding rocket experiment to measure Lyman emission from the resulting H-like C$^{5+}$ ion.  A schematic of the experimental apparatus can be seen in Fig.~\ref{detectorfig} and has been discussed previously \cite{defay2013}. The ion beam of $^{13}$C$^{6+}$ was produced by an electron cyclotron resonance (ECR) ion source with $^{13}$CO as the working gas. Isotopic carbon was selected to avoid ion beam contamination from similar mass-to-charge ratios. The $^{13}$C$^{6+}$ ion beam was extracted from the ECR ion source at 17.75 kV and momentum analyzed by a 90$^\circ$ dipole magnet. The ion source and analyzing magnet beam line are situated on a variable potential platform which can be operated in acceleration (positive potential) or deceleration (negative potential) mode to achieve the desired final ion beam energies. For this work, the final ion beam energies ranged from 1 -- 30 keV/amu (400 -- 2300 km/s).

Approximately 10 -- 30 nA of C$^{6+}$ ions were incident on a gas cell interaction volume (20 cm long) as shown in Fig.~\ref{detectorfig}.  The H$_2$ target gas was introduced into the gas cell volume via a leak valve with the total pressure being monitored via a nude Bayard-Alpert ion gauge and a quadrapole residual gas analyzer (SRS RGA100). The background pressure in the cell was $\simeq 10^{-8}$ Pa ($\simeq 10^{-10}$ Torr). The gas cell was held at a total pressure of $\simeq 10^{-6}$ Pa ($\simeq 10^{-8}$ Torr) during data acquisition. Due to the thermal recovery time of the detector between events, the pressure in the gas cell was adjusted slightly to restrict the x-ray count rate of the detector to $<1$~Hz for given ion beam currents.

The X-ray Quantum Calorimeter (XQC) detector has been described in detail elsewhere \cite{mccammon2002, mccammon2008} so only a brief overview will be given here.  The XQC is a 6 x 6 array of microcalorimeters with HgTe absorbers each 2.0mm x 2.0mm x 0.7$\mu$m. This array is situated in conjunction to an adiabatic demagnetization refrigerator resulting in a final operating temperature of 50~mK. The XQC was mounted to the interaction gas cell at 90$^\circ$ with respect to ion beam propagation, at a distance of 23~cm from beam line center. Because of the finite size of the detector array and the physical mounting limitations, only a limited portion (2 cm) of the gas cell was viewable by the detector. This is shown schematically by the viewing cone in Fig.~\ref{detectorfig}. The ions passed through this limited viewing distance in 10 -- 50~ns, for the given range of velocities investigated. This allowed for detection of prompt x-rays only due to charge exchange.

The effect of polarization due to capture to different $m$-states is not accounted for in our analysis. An isotropic emission is assumed, but it has been shown by others with a similar 90$^\circ$ detector orientation that this can lead to a maximum error (for fully polarized emission) of no more than 30\% and that typically less than 15\% is observed \cite{dijkkamp1985, hoekstra1989, greenwood2001}. 

\begin{figure}
\includegraphics[scale=0.5]{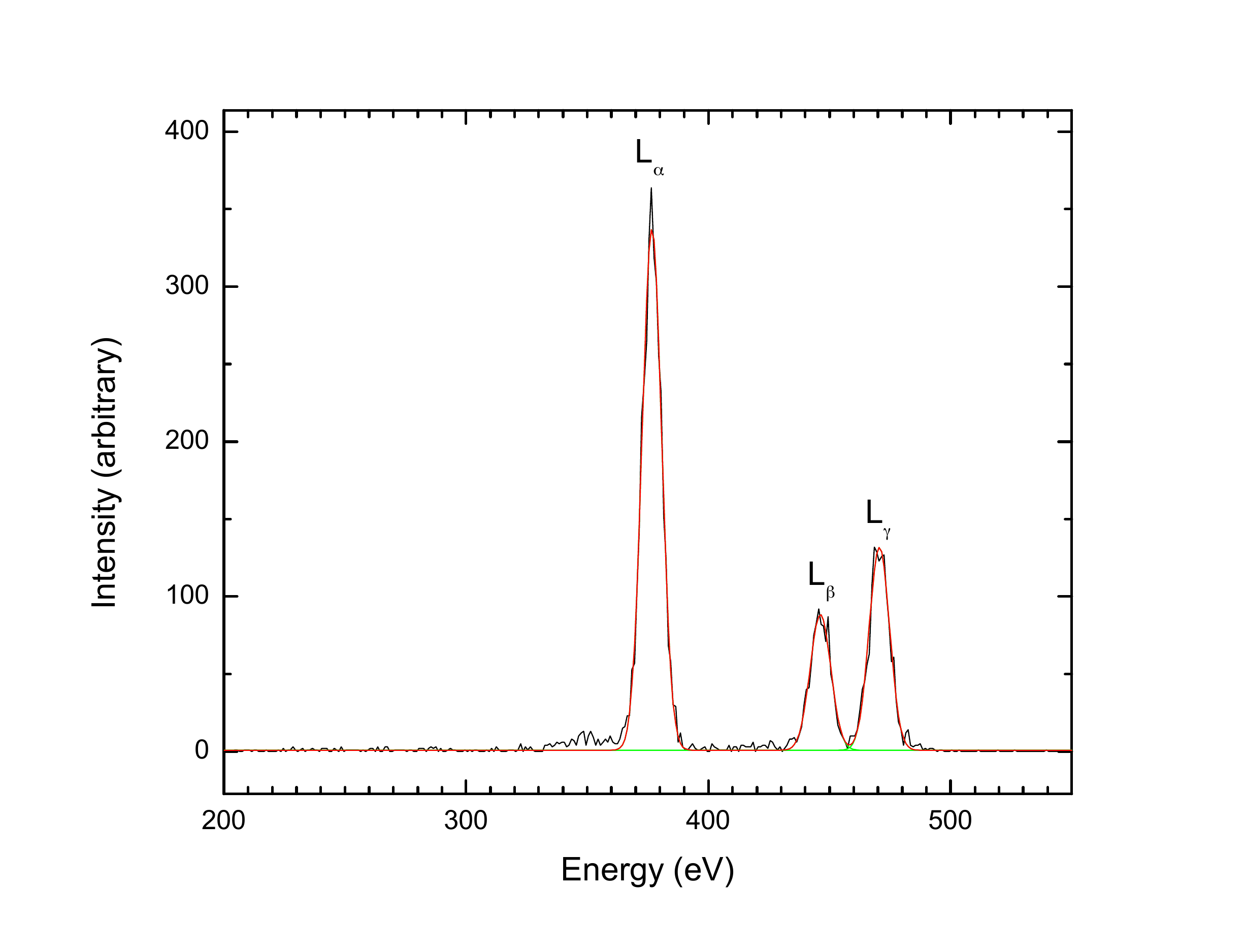}%
\caption{(color online) Representative Lyman series line spectrum and peak fit results from charge exchange of C$^{6+}$ and H$_2$ at 400 km/s. The emission line intensities have not been corrected for filter transmission.\label{linespectrum}}
\end{figure}

Figure~\ref{linespectrum} shows a typical x-ray spectrum recorded by the XQC for C$^{6+}$ on H$_2$ at a collision velocity of 400~km/s. Similar spectra were recorded over the range of collision velocities and each of the observed Lyman emission lines was peak fitted to obtain the integrated intensity of each emission line. A background spectrum, with no target gas in the gas cell, was taken at each energy to verify that there was no observable charge exchange contribution from the base residual gas of the gas cell. As can been seen in Fig.~\ref{linespectrum}, the full width half maximum (FWHM) line resolution of the XQC is approximately 10~eV. Because of various filters built into the XQC, the fitted line intensities must be corrected to account for energy-dependent transmission. This was done before final Lyman line ratios were determined. The net detection efficiencies for L$_\alpha$, L$_\beta$ and L$_\gamma$ lines are 0.065, 0.124 and 0.153, respectively.

\section{Results and Discussion}
From Fig.~\ref{linespectrum}, and the emission spectra recorded at other interaction velocities, we observed that only the L$_\alpha$, L$_\beta$ and L$_\gamma$ emission lines contributed due to charge exchange. There was no observation of Lyman-$\delta$ or higher lines, which suggests that only principle quantum states up to $n=4$ significantly contributed, in agreement with the CBM and the experimental VUV results of Dijkkamp et al. \cite{dijkkamp1985}. The results of Hoekstra et al. \cite{hoekstra1989} suggest a small ($\approx 3 $\%) contribution from capture to $n=5$.

\begin{figure}
\includegraphics[scale=0.5]{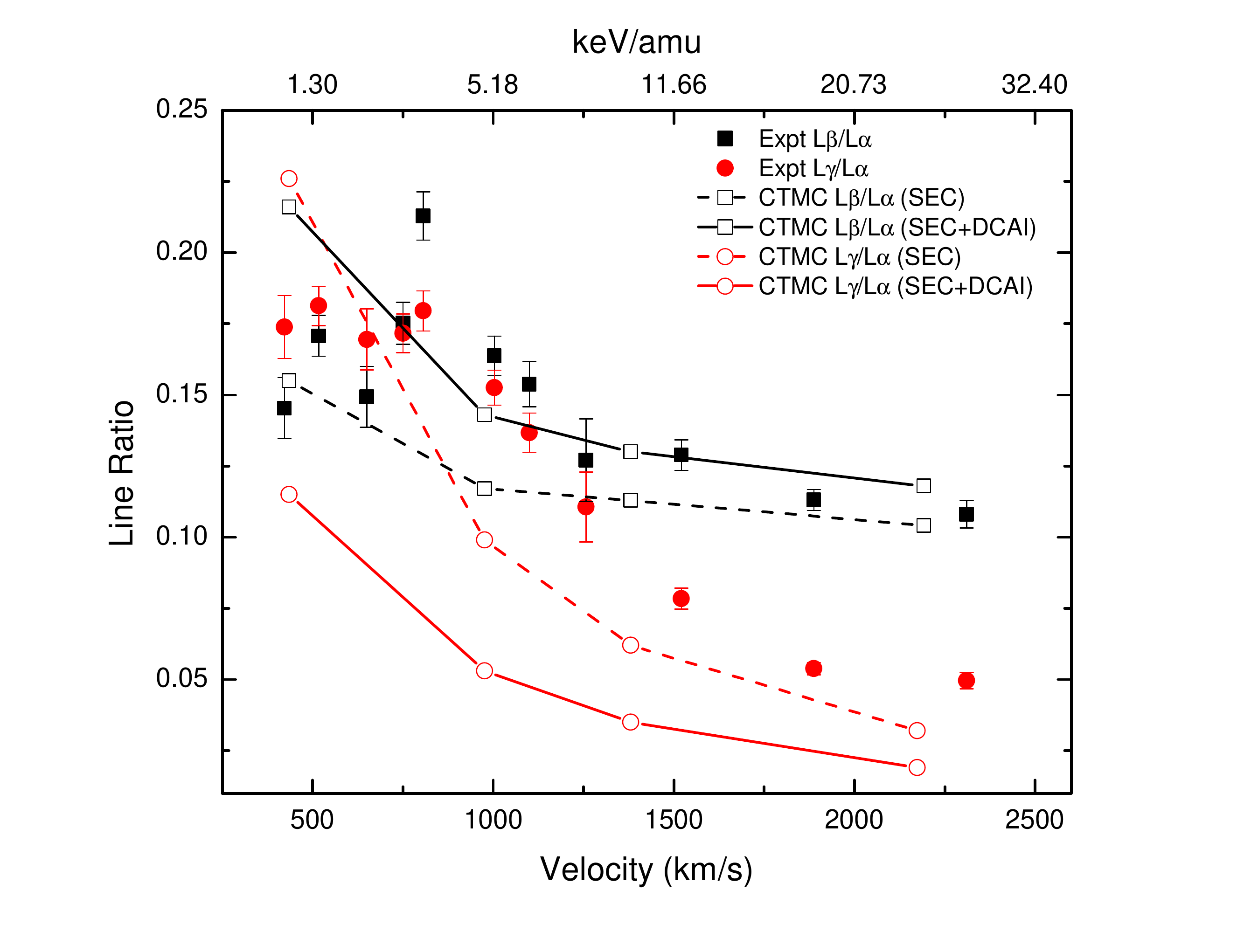}%
\caption{(color online) Measured L$_\beta$/L$_\alpha$ (filled squares) and L$_\gamma$/L$_\alpha$ (filled circles) line ratios as a function of collision velocity. Error bars represent 2$\sigma$ uncertainty. The open symbols represent CTMC calculations for SEC only (dashed joining lines) and SEC + DCAI (solid joining lines) \label{lineratioplot}}
\end{figure}

Figure~\ref{lineratioplot} shows the line emission ratios of L$_\beta$/L$_\alpha$ and L$_\gamma$/L$_\alpha$ as a function of collision velocity determined from the measured x-ray spectra. The L$_\beta$/L$_\alpha$ line ratios show a slight oscillatory behavior below 800~km/s with a peaked line ratio at 800~km/s. It is unclear how this oscillation and peaking behavior is connected to a physical phenomenon without guidance from a more rigorous theoretical investigation. It is more likely caused by a systematic uncertainty that is somehow not accounted for in the experiment. Above 800~km/s there seems to be a decreasing trend in the L$_\beta$/L$_\alpha$ line ratio with respect to increasing collision velocity.

The L$_\gamma$/L$_\alpha$ line ratios shown in Fig.~\ref{lineratioplot} exhibit fairly constant values below 800~km/s and clearly decrease with increasing collision velocity. At the highest measured velocity the L$_\gamma$/L$_\alpha$ line ratio is significantly less than the L$_\beta$/L$_\alpha$ line ratio although they had comparable values at the lower collision velocities.

If we consider only populating $n=$ 2, 3 or 4 in CX, and the resulting radiative cascades, the relative probabilities of producing L$_\alpha$, L$_\beta$ and L$_\gamma$ line emission due to CX to C$^{6+}$ can be determined by
\begin{eqnarray}
P(L_\alpha)&=&\frac{\sigma_2}{\Sigma \sigma_i} s_{2,1} + \frac{\sigma_3}{\Sigma \sigma_i}\left( s_{3,0}+s_{3,2} \right) \nonumber \\
&& + \frac{\sigma_4}{\Sigma \sigma_i}\left( s_{4,0}\frac{A_{4,0 \rightarrow 2,1}}{\Sigma A_{4,0}} + s_{4,1} \frac{A_{4,1 \rightarrow 3,0}+A_{4,1 \rightarrow 3,2}}{\Sigma A_{4,1}} + s_{4,2}\frac{A_{4,2 \rightarrow 2,1}}{\Sigma A_{4,2}} + s_{4,3} \right)\\
P(L_\beta)&=&\frac{\sigma_3}{\Sigma \sigma_i} s_{3,1} \frac{A_{3,1 \rightarrow 1,0}}{\Sigma A_{3,1}} + \frac{\sigma_4}{\Sigma \sigma_i} \left( s_{4,0} \frac{A_{4,0 \rightarrow 3,1}}{\Sigma A_{4,0}} \frac{A_{3,1 \rightarrow 1,0}}{\Sigma A_{3,1}} + s_{4,2} \frac{A_{4,2 \rightarrow 3,1}}{\Sigma A_{4,2}} \frac{A_{3,1 \rightarrow 1,0}}{\Sigma A_{3,1}} \right)\\
P(L_\gamma)&=& \frac{\sigma_4}{\Sigma \sigma_i} s_{4,1}\frac{A_{4,1 \rightarrow 1,0}}{\Sigma A_{4,1}}
\end{eqnarray}
where $\frac{\sigma_n}{\Sigma \sigma_i}$ is the relative cross section for an electron being captured to principle quantum state $n$. The relative weighting factor of each $n,l$ state is given by s$_{n,l}$ (the relative $l$-distribution normalized for each $n$), radiative transition A-values from state $n,l$ to $n',l'$ are given as $A_{n,l \rightarrow n',l'}$ (these values are given in Table I) and the total decay rate via all paths from a given $n,l$ state is given by $\Sigma A_{n,l}$.

\begin{table}
\begin{center}
\caption{A-values for the radiative transitions from $n,l$ to $n',l'$. Data statistically averaged over $j$ states were obtained from http://open.adas.ac.uk.}
\begin{tabular}{ccc}
\tableline \tableline
($n,l$) & ($n',l'$) & A$_{n,l \rightarrow n',l'}$ [s$^{-1}$]\\
\tableline
(2,1) & (1,0) & $8.12 \times 10^{11}$\\
(3,0) & (2,1) & $8.19 \times 10^9$\\
(3,1) & (1,0) & $2.17 \times 10^{11}$\\
(3,1) & (2,0) & $2.91 \times 10^{10}$\\
(3,2) & (2,1) & $8.38 \times 10^{10}$\\
(4,0) & (2,1) & $3.34 \times 10^9$\\
(4,0) & (3,1) & $2.38 \times 10^9$\\
(4,1) & (1,0) & $8.84 \times 10^{10}$\\
(4,1) & (2,0) & $1.25 \times 10^{10}$\\
(4,1) & (3,0) & $3.98 \times 10^9$\\
(4,1) & (3,2) & $4.51 \times 10^8$\\
(4,2) & (2,1) & $2.67 \times 10^{10}$\\
(4,2) & (3,1) & $9.12 \times 10^9$\\
(4,3) & (3,2) & $1.79 \times 10^{10}$\\
\tableline
\end{tabular}
\end{center}
\label{tab_Avalues}
\end{table}

From these calculated emission probabilities, the line ratios L$_\beta$/L$_\alpha$ and L$_\gamma$/L$_\alpha$ can then be determined for specific $s_{n,l}$ weighting factors and relative cross sections for CX resulting in an electron in $n=$ 2, 3 or 4. These can then be compared to the line ratios shown in Fig.~\ref{lineratioplot} to gauge the $l$-distribution model that best describes the Lyman line emission at different collision velocities.

Based on the CBM predictions, the majority of SEC is to $n=4$. In their VUV emission measurements of CX between C$^{6+}$ and H$_2$, Dijkkamp et al. \cite{dijkkamp1985} showed that approximately 90\% of the total capture cross section was to $n=4$ with an upper limit of approximately 10\% for capture to $n=3$. Hoekstra et al. \cite{hoekstra1989} suggested that 67\% of SEC capture is to $n=4$ and 6\% to $n=3$ with 3\% to $n=5$ and approximately 25\% to DCAI.

\begin{figure}
\includegraphics[scale=0.5]{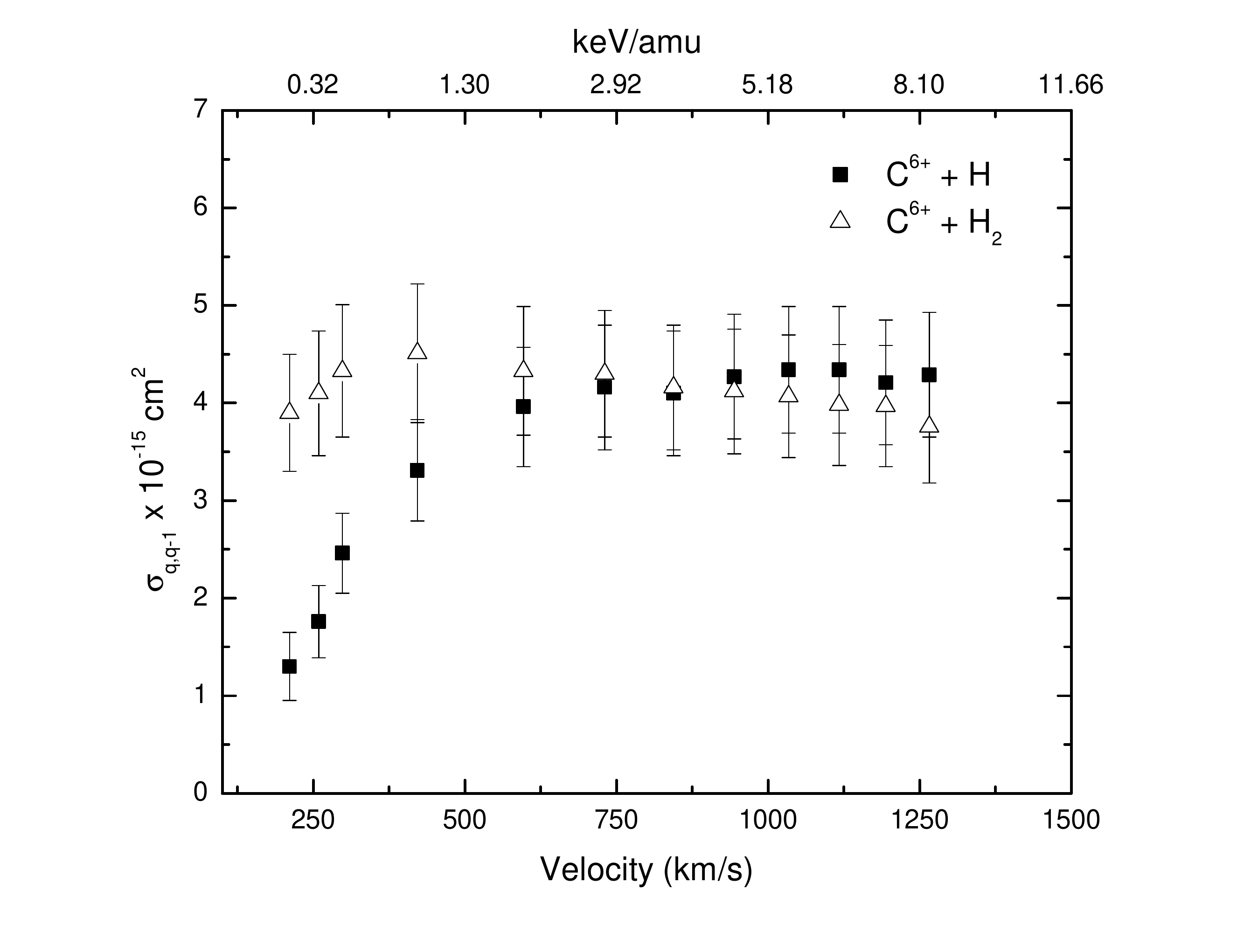}
\caption{Experimental total charge exchange cross sections for C$^{6+}$ on H and H$_2$ measured by Meyer et al. \cite{meyer1985}. \label{totalcsplot}}
\end{figure}

In order to gauge the possible contribution of DCAI at the lower collision velocities, we look to the total CX cross sections between C$^{6+}$ on H$_2$ and H measured by Meyer et al. \cite{meyer1985}. As can been seen in Fig.~\ref{totalcsplot}, the absolute CX cross sections from H$_2$ and H are essentially the same above 500~km/s, considering the uncertainty. However, slower collision velocities show a significant decrease in the CX cross section from H while the cross section from H$_2$ remains relatively unchanged. Given that these measurements could not distinguish SEC and DCAI, it is likely that DCAI is a significant contributor to the cross section difference observed at the slower velocities. 

In corresponding to the low end of our collision energy range, approximately 1.30~kev/amu, we determine the ratio of absolute cross section measurements of CX against H and H$_2$ from the data of Meyer et al. to be H:H$_2$=$0.82\pm0.18$. This suggests a DCAI contribution of $0.18\pm0.18$. The overall trend from the data of Meyer et al. suggests that the DCAI contribution decreases significantly with increasing collision energy.

Mack et al. \cite{mack1987,mack1989} have shown that the principal double capture states are of the form $3lnl'$ and $4lnl'$ with $3lnl'$ being significantly more dominant which results in an electron in a $2l$ state. In terms of the Lyman emission measurements presented here, this would lead to a relative increase of L$_\alpha$ for those DCAI events resulting in population of the $2p$ state.

\begin{figure}
\includegraphics[scale=0.6]{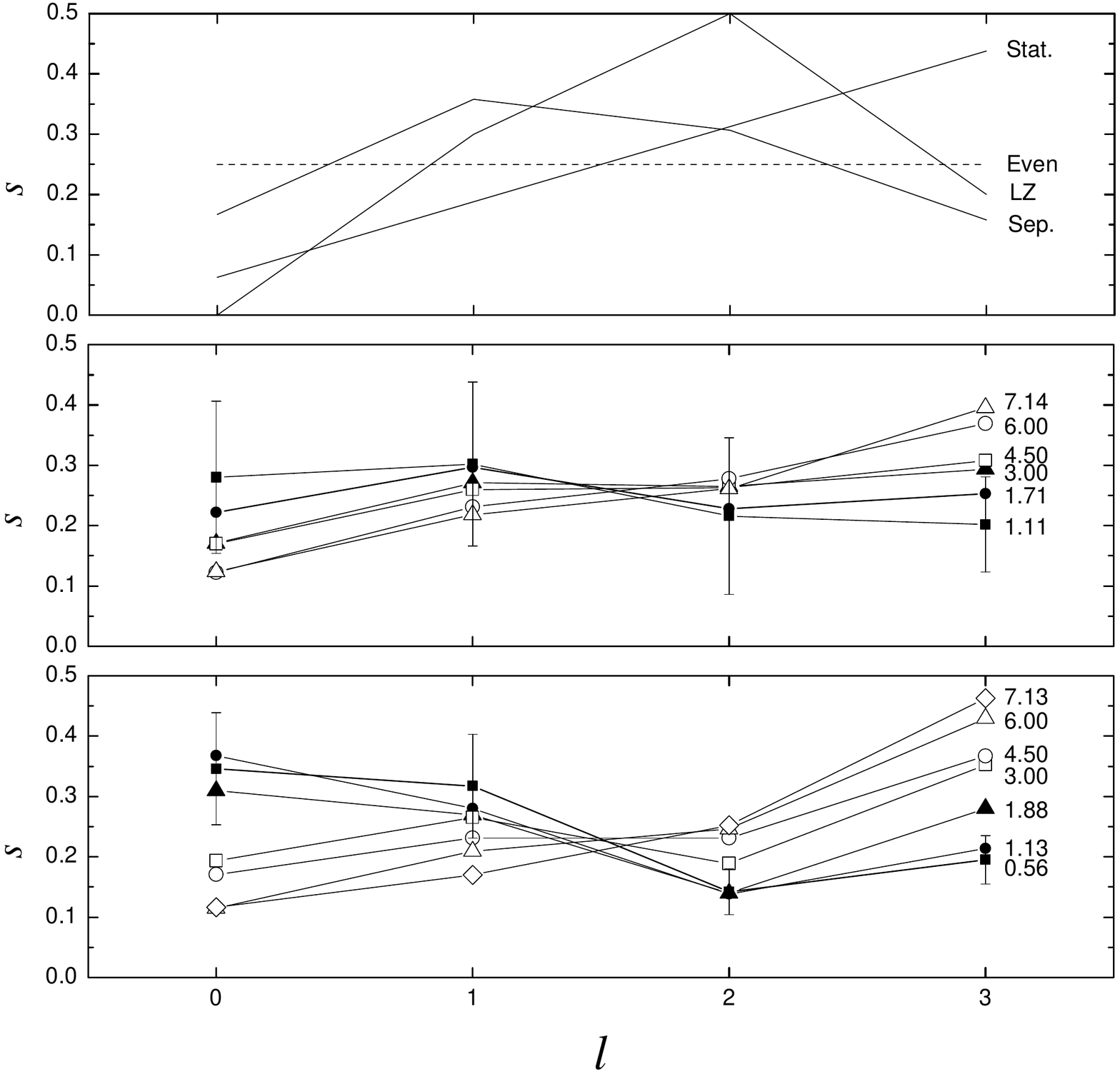}%
\caption{The upper pane shows the different $l$-distribution models for $n=4$. The middle pane shows the relative weighting factors derived from the state-selective CX cross sections of Dijkkamp et al. \cite{dijkkamp1985} for N$^{6+}$ on H$_2$ at various energies (keV/amu). The bottom pane shows the relative weighting factors derived from the state-selective CX cross sections of Dijkkamp et al. for O$^{6+}$ on H$_2$ at various energies (keV/amu). The relative level of uncertainty in the data of Dijkkamp et al. is reflected by the error bars attached to the lowest collision energy of each data set.\label{dijkkamp_data}}
\end{figure}

The upper pane of Fig.~\ref{dijkkamp_data} illustrates the relative $n=4$ weighting factors, s$_{4,l}$, for the  different $l$-distribution models discussed in the Introduction. In Table~II, we show the line ratios determined by substituting these $l$-distribution models into Eqns.~4--6 and considering only SEC to $n=4$. As can been seen, none of these calculated line ratios agree with those shown in Fig.~\ref{lineratioplot} over the range of velocities measured. This suggests that sole SEC to $n=4$ is not an adequate description of the overall CX and that some combination of capture to $n=2$ and 3 is necessary.

\begin{table}
\begin{center}
\caption{Calculated Lyman line ratios resulting from SEC to n=4 using different $l$-distribution models are shown on the first four rows. The last two rows show the calculated Lyman line ratios using the normalized n=4 $l$-distributions from Fritsch \& Lin \cite{fritsch1984} for C$^{6+}$ + H at 1~keV/amu and 25~keV/amu.}
\begin{tabular}{ccc}
\tableline \tableline
 $l$-distribution Model & L$_\beta$/L$_\alpha$ & L$_\gamma$/L$_\alpha$\\
\tableline
Even & 0.249 & 0.355 \\
Statistical & 0.130 & 0.221 \\
Landau-Zener & 0.192 & 0.431  \\
Separable & 0.261 & 0.603 \\
C$^{6+}$+H (1~keV/amu) & 0.202 & 0.350\\
C$^{6+}$+H (25~keV/amu) & 0.070 & 0.101\\
\tableline
\end{tabular}
\end{center}
\label{tab_ratios}
\end{table}

In evaluating the $l$-distribution model most applicable at the low collision velocities, we consider the VUV measurements of Dijkkamp et al. \cite{dijkkamp1985}. The middle and lower panes of Fig.~\ref{dijkkamp_data} show the relative weighting factors of the different $l$-states in $n=4$ for N$^{6+}$ on H$_2$ and O$^{6+}$ on H$_2$, respectively, at various collision energies. Dijkkamp et al. measured the $n,l$ cross section directly from their VUV line emission observations. The total cross section is derived by summing these state-specific cross sections. Dijkkamp et al. report uncertainties ranging from 20--50\% for various $n,l$ cross section measurements. These naturally carry over into the total cross section uncertainty. In determining the weighting factors shown in Fig.~\ref{dijkkamp_data}, the ratio of the state-specific cross section to the total cross section was taken. The resulting uncertainties were then weighted accordingly and added in quadrature to determine the relative uncertainties shown in Fig.~\ref{dijkkamp_data} attached to the lowest energy data set for each ion. In displaying just one of the error bars sets for each ion it is the intention to convey the relative uncertainty inherent to their data across the different collision energies.

In comparing the $l$-distributions with the model distributions in the upper pane of  Fig.~\ref{dijkkamp_data}, it can be seen that the relative uncertainties prevent any clear indication of which $l$-distribution model is most appropriate, however, the data for O$^{6+}$ seems to show a trend of transitioning from a low-$l$ distribution at the lowest collision energies towards a more statistical distribution at higher collision energies. Given the uncertainty, there is clearly an increase in the population of $l=3$ as the collision velocity increases which indicates that the lower $l$-state populations must be shifting to higher $l$, as the overall trend seems to suggest. 

The N$^{6+}$ data, given the uncertainty, seems to be more reflective of an even $l$-distribution across the collision energies with perhaps a slight transition towards a more statistical distribution at higher collision energies. This, along with the O$^{6+}$ data, is perhaps suggesting a trend towards a more even/statistical $l$-distribution for C$^{6+}$ at the lower end of our observed collision energies. The average of our observed line ratios for L$_\beta$/L$_\alpha$ and L$_\gamma$/L$_\alpha$ at lower collision velocities and their disparity with the different $l$-distribution models for SEC to $n=4$ only, as shown in Table~II, suggests that we do not observe the amount of CX to $3p$ and $4p$ states predicted by $l$-distributions peaked at low $l$-states, such as in the LZ and Separable $l$-distribution models. The overall trend of the Dijkkamp et al. data seems to suggest this as well. This points to the statistical $l$-distribution model as the appropriate in this collision velocity regime. It should also be noted that the collision velocity in this range corresponds with approximately half the orbital velocity of the captured electron, which is typically considered to result in a statistical $l$-distribution\cite{burgdorfer1986}.

\begin{figure}
\includegraphics[scale=0.75]{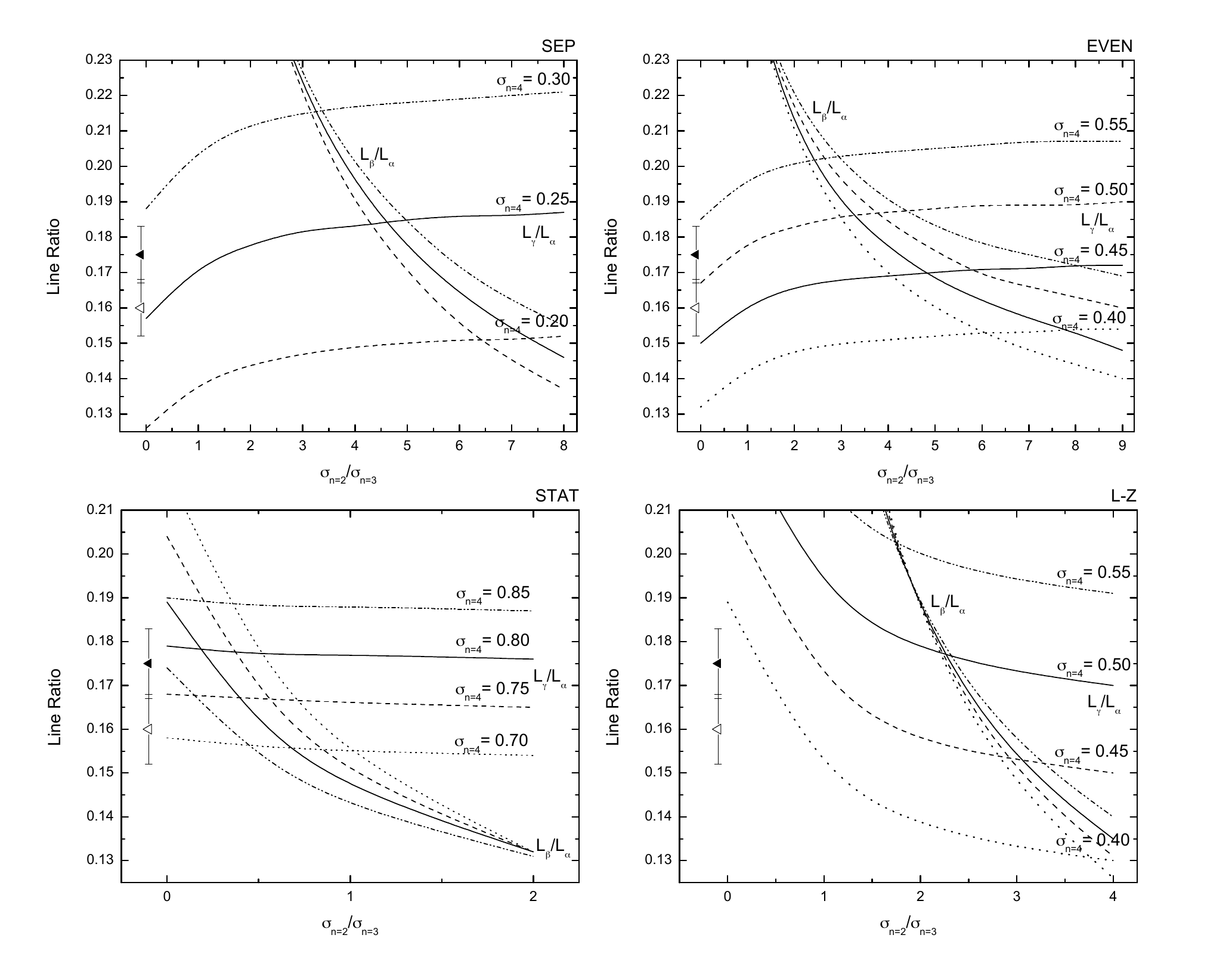}
\caption{Effect of relative cross section partitioning to line ratios for different $l$-distribution models. The line types are paired based on the relative $n=4$ cross sections.  The filled symbol on the left of each plot represents the mean L$_\gamma$/L$_\alpha$ line ratio below 800~km/s. The open symbol represents the mean L$_\beta$/L$_\alpha$ line ratio below 800~km/s. \label{npartplot} }
\end{figure}

To further illustrate the effect that the different $l$-distribution models have on the L$_\beta$/L$_\alpha$ and L$_\gamma$/L$_\alpha$ line ratios, we can now consider the relative cross sections to $n=$ 3 and 4 via SEC and a possible contribution to $n=2$ via DCAI at low collision velocities by comparing to calculated line ratios using the various $l$-distribution models from the Introduction and comparing to our observed mean line ratios below a collision velocity of 800~km/s.

Each pane of Fig.~\ref{npartplot} illustrates two forms of partitioning simultaneously for each $l$-distribution model. Each pair of linetypes in the various plots represents the calculated (from Eqns. 4--6) L$_\beta$/L$_\alpha$ and L$_\gamma$/L$_\alpha$ line ratios for a fixed relative cross section for population of $n=4$ (primary SEC capture state). The remaining relative cross section contribution is then distributed between $n=2$ and 3 as shown by the ratio of $\sigma_{n=2}/\sigma_{n=3}$ in each plot. We have selected the different relative cross sections for $n=4$ in discrete steps to match the range of our observed line ratios and to give an idea of the overall behavior. Values can be interpolated as needed.

From the separable $l$-distribution data in Fig.~\ref{npartplot} it can be seen that to achieve line ratios consistent with the observed L$_\beta$/L$_\alpha$ and L$_\gamma$/L$_\alpha$ line ratios, a relative $n=4$ cross section of 0.23--0.24 would be required along with a $\sigma_{n=2}/\sigma_{n=3}$ of 5--6, suggesting a relative $n=2$ contribution, from DCAI, on the order of 0.60 and a SEC contribution to $n=3$ of approximately 0.17. Based on the absolute cross section data of Meyer et al. this model seems to suggest a considerably smaller $n=4$ relative cross section and a considerably larger DCAI contribution. This DCAI contribution is also larger than the upper limit recommendation of Hoekstra et al. (25\%) and seems to be at odds with the overall trend in the data of Meyer et al. at the given collision velocity.

The even $l$-distribution data in Fig.~\ref{npartplot} shows that a relative $n=4$ cross section of 0.45--0.47 along with $\sigma_{n=2}/\sigma_{n=3}$ of 5--8 yields line ratios comparable to our observed line ratios at low collision velocities. This corresponds to a relative DCAI contribution of 0.45--48. This is also considerably more than is suggested from the absolute measurement of Meyer et al. and the upper limit suggested by Hoekstra et al. This, however, is one of the likely $l$-distributions that seems to agree with the weighting factors derived from the data of Dijkkamp et al. shown in Fig.~\ref{dijkkamp_data}.

The LZ $l$-distribution data in Fig.~\ref{npartplot} shows that a relative $n=4$ cross section of 0.50 and a corresponding $\sigma_{n=2}/\sigma_{n=3}$ of 2.5--3.0 agrees well with the observed line ratios. This suggests a relative DCAI contribution of approximately 0.36 which is just within the uncertainty range for the DCAI contribution determined from the data of Meyer et al. but is still larger than the upper limit suggested by Hoekstra et al.

The statistical $l$-distribution data in Fig.~\ref{npartplot} shows that a relative $n=4$ cross section of approximately 0.80 and a correspond $\sigma_{n=2}/\sigma_{n=3}$ 0.4--0.6 agrees well with the observed line ratios. This suggests a relative DCAI contribution of approximately 0.07 and a relative $n=3$ contribution of 0.13. The relative $n=3$ contribution is in agreement with that of Dijkkamp et al. (10\%) and is twice that suggested by Hoekstra et al. (6\%). The relative DCAI contribution ($n=2$) is considerably less than the suggested upper limit of 25\% by Hoekstra et al. but seems to be in agreement with the absolute cross section data of Meyer et al. We would also like to point out that the statistical $l$-distribution is also one of the likely suggested models that corresponds to the $l$-distribution data of Dijkkamp et al. In light of all of the results from these models in comparison to the absolute measurements of Meyer et al. and the experimental data of Dijkkamp et al. and Hoekstra et al. The statistical $l$-distribution seems the most likely model to describe our observed line ratios below 800~km/s. It would, however, be desirable for more complete theoretical models to verify if this is a proper description of the CX dynamics for this collision system.

If we assume a statistical $l$-distribution model to describe our observed line ratios at the low collision velocities, then we can assume that at higher collision velocities that higher $l$-state populations will increase leading to a more over-statistical distribution. Since we don't have a model function to guide an investigation of that description of the observed CX, we can compare to CX data for C$^{6+}$ on H. This is a single electron system but the absolute CX cross section data of Meyer et al. suggests that at higher collision velocities the cross sections are comparable and thus no considerable contribution from DCAI is expected. The absolute measurements will also have taken into account the small contribution of SEC to $n=3$ in the H$_2$ case.
\begin{figure}
\includegraphics[scale=0.6]{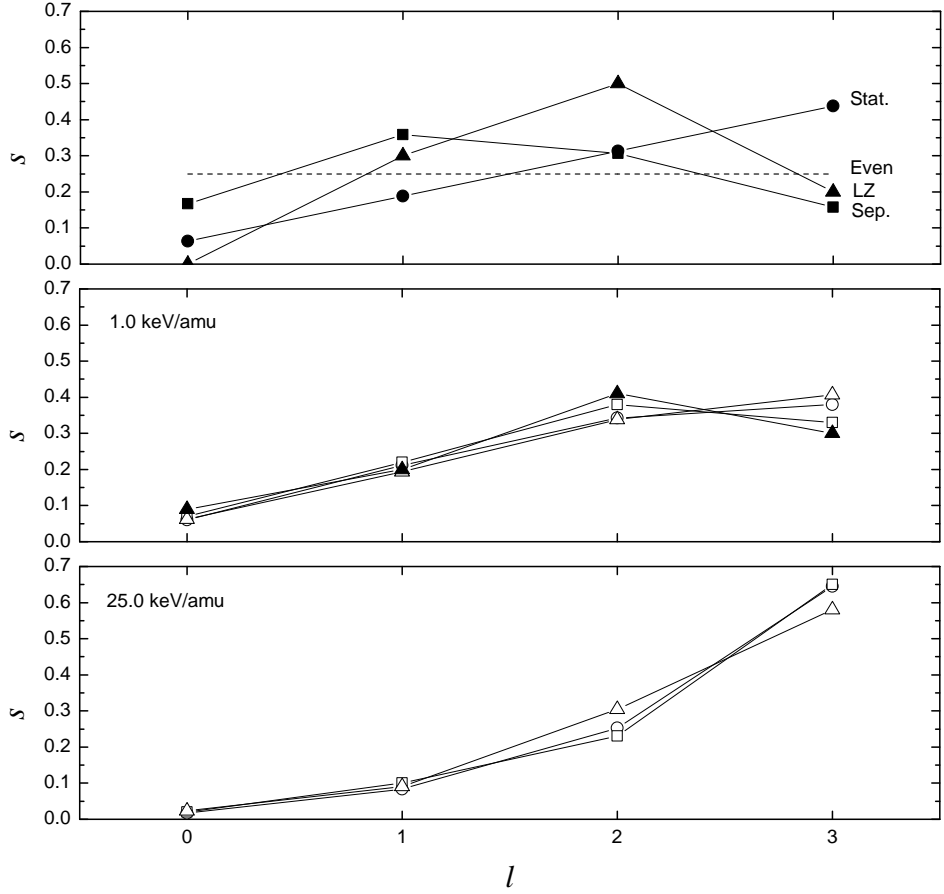}%
\caption{The upper pane is the same as from Fig.~\ref{dijkkamp_data} reproduced here for comparison. The middle and bottom panes show the relative weighting factors for $n=4$ derived from the state-selective CX cross sections of Toshima \& Tawara \cite{toshima1995} (open circles), Fritsch \& Lin \cite{fritsch1984} (open squares), Green et al. \cite{green1982} (open triangles) and Kimura \& Lin \cite{kimura1985} (filled triangles) for C$^{6+}$ on H at 1~keV/amu and 25~keV/amu, respectively. \label{H_data}}
\end{figure}

In the comparison of the resulting $l$-distributions for capture to $n=4$ in CX between C$^{6+}$ and H$_2$ and H at different energies, it is expected that the relative $n$-state populations will be slightly different given the ionization potential difference between H$_2$ and H. For the H target at 1~keV/amu, approximately 90\% of the total capture cross section is to $n=4$ while at 25~keV/amu, capture to $n=5$ becomes a substantial part of the total cross section.
Figure \ref{H_data} shows the relative $l$-distributions for the $n=4$ manifold from the data of Toshima \& Tawara \cite{toshima1995}, Fritsch \& Lin \cite{fritsch1984}, Green et al. \cite{green1982} and Kimura \& Lin \cite{kimura1985} at 1 keV/amu and 25 keV/amu collision energies. This range is comparable with our measured lower and upper energy range. As can bee seen in Fig.~\ref{H_data}, the H data at 1~keV/amu  suggests that the statistical and LZ $l$-distributions are most predominant while at 25 keV/amu, the $l$-distribution population is peaked at the maximum $l$-state, suggesting a transition to an over-statistical distribution. As a comparison, Table II shows the L$_\beta$/L$_\alpha$ and L$_\gamma$/L$_\alpha$ line ratios calculated using the n=4 $l$-distribution data of Fritsch \& Lin for capture from atomic hydrogen at 1~keV/amu and 25~keV/amu. At 1~keV/amu, the calculated line ratios are similar to the those obtained from the LZ model. At 25~keV/amu, the line ratios are closer to those observed for capture from H$_2$ and are indicative of the trend to capture to higher angular momentum states at higher collision energy. We note that at the higher collision energy, the predominant capture to maximum angular momentum leads to increased L$_\alpha$ emission and thus lower overall line ratios with respect to L$_\alpha$, as would be expected.
Finally, Figure~\ref{lineratioplot} shows the results of preliminary CTMC calculations for both the L$_\beta$/L$_\alpha$ and L$_\gamma$/L$_\alpha$ line ratios \cite{stancil2013, hasan2001}. As pointed out previously, the CTMC method is expected to be most applicable at the higher collision velocities. The calculations are done for the cases of SEC only and for SEC in combination with DCAI and TI, although the TI cross section contribution is an order of magnitude less than the DCAI contribution and is considered negligible. As can be seen in Fig.~\ref{lineratioplot} at the higher collision velocities, the CTMC results are in relatively good agreement with the line ratios determined from the experiment and tend to be independent of the DCAI contribution, which is reasonable given that DCAI is expected to fall off quickly with increasing collision velocity. At the low collision velocities, the CTMC results do not agree with either the L$_\beta$/L$_\alpha$ or L$_\gamma$/L$_\alpha$ line ratios. This is likely an indication that the applicability of the CTMC model in this collision range is not appropriate.

\section{Conclusion}
State-specific $n,l$ CX cross sections have been known to be highly dependent on collision velocity, however, it is still relatively difficult to predict emission spectra using the predominant $l$-distribution models put forth in the literature. From our measured high resolution x-ray emission spectra for CX between C$^{6+}$ and H$_2$, we observe comparable L$_\beta$/L$_\alpha$ and L$_\gamma$/L$_\alpha$ line ratios at low collision velocities that decrease with increasing collision velocity. 

Previous absolute cross section measurements and VUV line emission experimental results provide some guidance towards applicable $l$-distribution models but are not conclusive. In comparing our observed line ratios with calculated line ratios using the separable, even, Landau-Zener and statistical models we determine that the statistical model best describes line emission at the low end of our observed collision velocities. This implies a transition to a more over-statistical $l$-distribution at higher collision velocities. There is, however, an impelling case for more rigorous theoretical models to investigate this collision system over this velocity range to determine the range of applicability for different theoretical approaches.

\bibliography{myrefs}

\begin{acknowledgments}
This research is supported in part by the NASA Solar \& Heliospheric Physics Program NNH07ZDA001N, NASA Grant No. NNX09AF09G, and by the Office of Fusion Energy Sciences and the Division of Chemical Sciences, Geosciences and Biosciences, Office of Basic Energy Sciences, US Department of Energy. We would like to thank P.~C.~Stancil and D.~R.~Schultz for providing the CTMC calculation results. The authors would also like to extend a special thanks to S.~D.~Loch and C.~Ballance for stimulating discussions and suggestions. 
\end{acknowledgments}



\end{document}